# A New Embedded Measurement Structure for eDRAM Capacitor

[1&2]L. Lopez                [1]J.M. Portal                [2]D. Née

[1] L2MP-Polytech-UMR CNRS 6137
IMT - Technopôle de Château Gombert
F-13451 Marseille, France
portal@polytech.univ-mrs.fr
Tel:(33)-491-054-787

[2] ST-Microelectronics
ZI de Rousset BP 2
F-13106 Rousset, France
didier.nee@st.com
Tel:(33)-442-688-815

**Abstract**

*The embedded DRAM (eDRAM) is more and more used in System On Chip (SOC). The integration of the DRAM capacitor process into a logic process is challenging to get satisfactory yields. The specific process of DRAM capacitor and the low capacitance value (~30fF) of this device induce problems of process monitoring and failure analysis. We propose a new test structure to measure the capacitance value of each DRAM cell capacitor in a DRAM array. This concept has been validated by simulation on a 0.18μm eDRAM technology.*

## 1. Introduction

The major challenge of eDRAM technologies remains the significant increase in the fabrication process complexity with its associated control problems. To face this major challenge, Built-In-Self-Repair (BISR) techniques are extensively developed. Complementary to these BISR techniques, some failure analysis methods have been proposed [1,2], knowing that memories offer specific properties of regularity. But none of these methods targets the parametric values of eDRAM such as capacitor value. To enhance the failure analysis efficiency, it appears that embedded capacitor measurement function is a very relevant feature to have information on the specific eDRAM process steps. Thus, the fundamental discussion of this paper is to present an embedded capacitor measurement function for eDRAM, and the associated diagnosis methodology improvement.

## 2. Capacitor measurement structure

As illustrated Figure 1, the proposed measurement structure is connected to the plate node of the macro-cell (to simplify the scheme, only four cells of the macro-cell are represented) in order to delete capacitance noise measurement due to the parasitic bit-line capacitance. This structure is switch off in the standard operation mode and the plate polarization is fixed to $V_{DD}/2$ using the STD transistor. During the test mode, the STD transistor is switched off. The Select transistors allow the memory array to connect each bit-line to the bit-line input named $IN_{BLi}$. These transistors are named $S_{BLi}$.

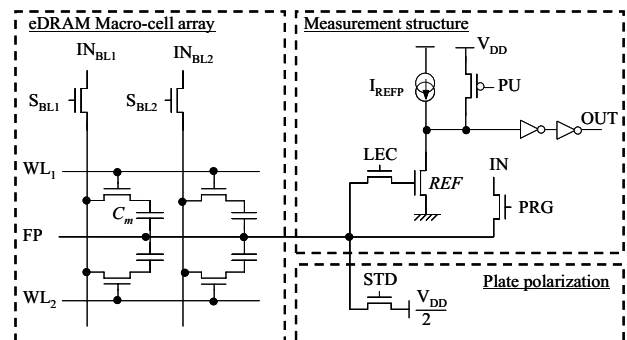

**Figure 1: eDRAM Macro-Cell with capacitor extraction structure connected to the plate node**

The measurement structure itself is composed of two select transistors controlled respectively by the signal LEC and PRG, one reference transistor REF with an input capacitor $C_{REF}$ and one programmable current reference named $I_{REFP}$. The sensing function is composed of two inverters, which drive the digital output OUT.

The measurement procedure is based on charge sharing between the capacitor under measurement ($C_m$) and the reference capacitor ($C_{REF}$). The reference capacitor is the input capacitor of the n-MOSFET used for the analog to digital conversion.

The measurement flow is composed of five steps of 10 ns. The first step is the discharge off all the capacitor in the macro-cell and in the measurement structure. To do so, all the world lines are selected (DRAM transistors ON). The n-MOSFET controlled by LEC is also turned on. All the capacitors are then grounded on their both nodes: PRG is conducting and by applying 0V on the input IN, the plate node is grounded; All $S_{BLi}$ are conducting and by applying 0V to each bitline inputs ($IN_{BLi}$ are grounded), all the bit line nodes are grounded.

The second step of the measurement flow consists in charging the capacitor $C_m$ that is controlled by $WL_1$, without charging all the others capacitors in the macro-cell. This operation is performed by turning off all the word lines except $WL_1$, and by raising all the bit-lines to $V_{DD}$ except the one that is connected to $C_m$ that remains grounded. LEC is grounded to unselect $C_{REF}$ and applying $V_{DD}$ on the input IN performs the charge of $C_m$. At the end of the step, PRG is turned off.

The third step is straightforward and consists in turning off the bit-line selection signal $S_{BLi}$ except the one of $C_m$ to put on high impedance state all the other capacitors of the macro-cell. In this configuration, the capacitor $C_m$ is the only one still active on the plate node.

The fourth step consists in turning on the LEC signal. A charge sharing is made between $C_m$ and $C_{REF}$. Thus, the potential $V_{GS}$ is a function of $C_m$.

The last step of the measurement flow is to get a numerical data of $V_{GS}$ and thus, of $C_m$. This step is based on a current injection through the transistor REF. The programmable current source $I_{REFP}$ has been designed to get a numerical linear ramp of current with 20 steps controlled by a shift register [3]. The current rises by step and the drain potential rises too. When $V_{DS}$ is larger than $V_{DD}/2$, the inverter connected to the drain of REF switches from a logical 1 to a logical 0. Thus, the output (OUT) switches from a logical 0 to a logical 1. The stored value in the shift register that is used to control the current ramp, is then extracted on the output switches, and gives a digital image of the capacitor's value.

This structure has been validated by simulation using a design kit of the 0.18μm eDRAM technology from ST-Microelectronics. The Figure 2 presents simulation results with $C_m$ set to 30fF. The capacitance variation induces a variation of time of the OUT switch and so, a variation of the current steps of $I_{REFP}$. The conversion of the register value to the image of the capacitor one is straightforward. The switch of the output OUT occurs for different current steps depending on the capacitor value. The register value gives directly the current step. This current value is used as an image of the capacitor value, thus a specification window is defined in current. Using the abacus obtained from a set of simulation, Figure 3 shows the current steps versus the capacitor values. With our design, the test structure is scaled in a range of eDRAM capacitor of 10fF-55fF with an accuracy of 6%. If the number of current step is 0, three diagnoses are possible: The capacitor value is under 10fF; the capacitor is shorted; the capacitor behaves like an open. If the number of current step is 20, the capacitor value is equal or superior to 55fF. The main idea, when extracting the capacitor value, is to build an Analog Bitmap of the capacitor values of the cells in the memory array. This analog bitmap can be treated in the same way than the digital one, with signatures categorization depending on the capacitor values. This signatures categorization might be very useful to characterize process and defect impact on the array.

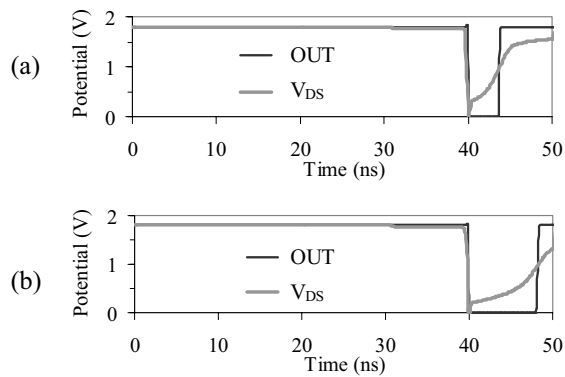

**Figure 2: Capacitor extraction simulation results : (a) Cm = 20 fF; (b)Cm = 40 fF)**

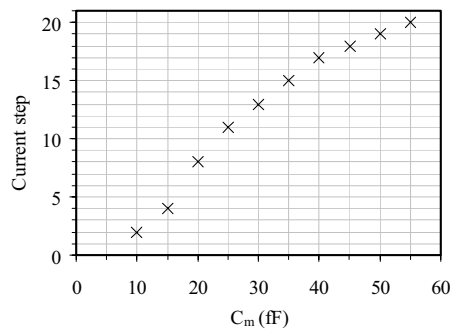

**Figure 3: Abacus to define the equivalence between current step and capacitor value.**

## 3. Conclusion

In this paper, an embedded test structure is developed to extract the capacitor values of each cell in the array during the functional test. The capacitor values are extracted in a digital format that enables a diagnosis methodology based on analog bitmapping complementary to the classical digital bitmapping. Thus, the diagnosis of failure of each cell in the array is improved.